\documentclass[pra,twocolumn,showpacs]{revtex4-1}
\usepackage{graphicx}
\usepackage{amsmath}
\usepackage{amssymb}
\usepackage{amsfonts}
\usepackage{mathrsfs}
\usepackage{xspace}
\usepackage{verbatim}
\usepackage{graphicx}
\usepackage{esint}
\usepackage{hyperref}
\usepackage{epstopdf}
\usepackage{nicefrac}
\usepackage{verbatim}
\usepackage[version=4]{mhchem}
\usepackage[normalem]{ulem}
\usepackage{inputenc}
\usepackage{siunitx}
\begin{document}

\title{Topological defect dynamics of vortex lattices in Bose--Einstein condensates}
\author{Lee~James~O'Riordan and Thomas~Busch}
\affiliation{Quantum Systems Unit, Okinawa Institute of Science and Technology Graduate University, Onna, Okinawa 904-0495, Japan.}

\begin{abstract}
Vortex lattices in rapidly rotating Bose--Einstein condensates are systems of topological excitations that arrange themselves into periodic patterns. Here we show how phase-imprinting techniques can be used to create a controllable number of defects in these lattices and examine the resulting dynamics. Even though we describe our system using the mean-field Gross--Pitaevskii theory, the full range of many particle effects among the vortices can be studied. In particular we find the existence of localized vacancies that are quasi-stable over long periods of time, and characterize the effects on the background lattice through use of the orientational correlation function, and Delaunay triangulation.

\end{abstract}


% 03.75.-b	Matter waves
% 67.85.-d	Ultracold gases, trapped gases

\maketitle

\section{Introduction}\label{sec:intro}
%%%%%%%%%%%%%%%%%%%%%%%%%%%%%%%%%%%%%%%%%%%%%%%%%%%%%%%%%%%%%%%%%%%%%%%%%%%%%%%%%%%%%%%%%%%%%%%%%%%%%%%%%%%%%%%%%%%%%%%%%%%%%%%%%%%%%%%%%%%%
Ultracold gases have proven to be a valuable resource for building simulators of condensed matter and solid-state systems \cite{Bloch:08,Flaschner_sci_2016,Suchet_epl_2016}. This is due to the fact that they are low energy systems that can be trapped in periodic settings using optical lattices, and that a large number of techniques exist to control and change all terms of the corresponding Hamiltonians. Control over the lattice depths, and therefore the tunneling strength, gives a handle on the kinetic energy term \cite{AO:Jaksch_prl_1998}, superlattices allow to adjust the on-site energies \cite{AO:Strabley_pra_2006}, and different external lattice geometries lead to changes in the band structure \cite{AO:Eckhardt_epl_2010,Tarruell:12}. Most recently artificial gauge fields have been added to the toolbox \cite{Hauke_prl_2012,Vtx:LeBlanc_njp_2015,Raventós_pra_2016}.

One of the most interesting topics in solid-state physics is the study of impurities and their effect on the background system. Studying impurities in Bose--Einstein condensates (BECs) is a promising and rich topic and the first experiments in this area have recently been carried out \cite{AO:Hohmann_epjqt_2015, BEC:Schmid_prl_2010, BEC:Zipkes_nat_2010}. Given the clean and highly controllable nature of condensates, they allow a study of the fundamental physics of impurities, which is paramount to creating models of realistic condensed matter systems, which are never truly impurity or defect-free. By now impurities have been used to investigate the atomic density distribution \cite{BEC:Schmid_prl_2010}, as well as exotic quasi-particles such as Fr\"ohlich polarons \cite{AO:Hohmann_epjqt_2015}. These results show that impurities are very robust and reliable tools to investigate the underlying condensate behaviour, and many proposals for further investigations exist \cite{CG:Goold_pra_2011,BEC:Ardila_pra_2015,Vtx:Johnson_prl_2016,BEC:Grusdt_pra_2016,CG:Cetina_arxiv_2016}.

Another ultracold system in which solid-state-like periodic structures appear are vortex lattices in rotating Bose--Einstein condensates. BECs react to high rotation frequencies by creating a large number of vortices with single winding, which arrange themselves into a triangular Abrikosov geometry, similar to the ones observed in type-II superconductors \cite{AO:Abrikosov_rmp_2004}. These lattices have been investigated for their collective excitations, and have been shown to exhibit Tkachenko mode behavior \cite{BEC:Coddington_prl_2003,BEC:Baym_prl_2003}. More recently the focus has shifted towards looking at perturbations of these lattices. For example, applying a kicked potential with a spatial geometry similar to the vortex lattice was shown to create transient superlattice structures in the density \cite{VTX:oriordan_pra_2016}. Quantifying the disorder of vortex lattices has recently become an active topic of interest \cite{Vtx:Mithun_pra_2016,Rakonjac:16}. These topics are particularly useful as they can allow the study of quantum turbulence in highly controllable systems \cite{Vtx:Neely_prl_2013,Vtx:Kwon_pra_2014,Vtx:Groszek_pra_2016}.

All the studies up to now have focussed on collective behavior of the vortex lattices, as the introduction of a single impurity into a vortex lattice is hard to achieve. In this work we suggest an experimentally realistic method to do this and examine the behavior of a vortex lattice in the presence of a defect or impurity. For this we start from a perfect vortex lattice at a fixed rotation frequency, and will selectively either remove single vortices, or introduce additional rotation at localized positions. The easiest and experimentally most realistic way to do this is via phase-imprinting, and we will show that this is a highly controllable and precise way to manipulate vorticity in a rapidly rotating Bose--Einstein condensate.

The manuscript is organized as follows. In Sec.~\ref{sec:model} we introduce the system of a rapidly rotating Bose--Einstein condensate featuring a vortex lattice. We then proceed to investigate the dynamics of removing a vortex from the condensate using the phase imprinting method in Sec.~\ref{sec:phase} and conclude in Sec.~\ref{sec:Conclusions}.

%%%%%%%%%%%%%%%%%%%%%%%%%%%%%%%%%%%%%%%%%%%%%%%%%%%%%%%%%%%%%%%%%%%%%%%%%%%%%%%%%%%%%%%%%%%%%%%%%%%%%%%%%%%%%%%%%%%%%%%%%%%%%%%%%%%%%%%%%%%%%
\section{Model}\label{sec:model}
%%%%%%%%%%%%%%%%%%%%%%%%%%%%%%%%%%%%%%%%%%%%%%%%%%%%%%%%%%%%%%%%%%%%%%%%%%%%%%%%%%%%%%%%%%%%%%%%%%%%%%%%%%%%%%%%%%%%%%%%%%%%%%%%%%%%%%%%%%%%%

For this work we consider the system of an Abrikosov vortex lattice in a rapidly rotating Bose--Einstein condensate within the mean-field regime.
To investigate the evolution of a perturbed vortex lattice we numerically solve the Gross--Pitaevskii equation in two dimensions, assuming a strong confinement along the third axis. This allows us to restrict the dynamics to the $x$--$y$ plane and focus fully on the Abrikosov lattice geometry. Experimentally, this corresponds to a system with a very strong confinement in one direction only \cite{BEC:Stock_lpl_2004,BEC:Seo_jkps_2014,BEC:Chomaz_natcom_2015} and in the frame co-rotating with the condensate, the non-linear mean field equation governing the BEC wave-function is given by
\begin{align}
	\mathrm{i}\hbar\partial_t \Psi(\mathbf{r},t) = \Big[&-\frac{\hbar^2}{2m}\nabla^2 + V\left(\mathbf{r}\right) \nonumber\\
	&+ g\vert \Psi(\mathbf{r},t) \vert^2- \Omega L_z \Big]\Psi(\mathbf{r},t).
\end{align}
Here $V\left(\mathbf{r}\right)$ is the harmonic trapping potential with a frequency $\omega_\perp=2\pi\times1$ Hz.  The trap rotation frequency is given by $\Omega$ and $L_z$ is the angular momentum operator along the $z$-direction. The effective interaction strength in 2D is characterized by $g$ and we assume to have $N=9.8\times 10^5$ atoms of $^{87}$Rb, with a singlet state \textit{s}-wave scattering length of $a_s = 90r_b \approx 4.76\times 10^{-9}$ m, where $r_B$ is the Bohr radius. For the rapidly rotating case, $\Omega=0.995\omega_\perp$, the vortices form an ordered triangular lattice with spacing $a_v \approx 2.1\times 10^{-5}$ m, that rotates similarly to a solid-body in the large number limit \cite{BEC:Fetter_rmp_2009}. Simulating a large vortex lattice is a difficult numerical problem, as large grid sizes are required to resolve all aspects of the system both in position and momentum space. Thus, an advanced numerical technique is necessary to obtain solutions in a reasonable timescale. We have developed and made use of ``GPUE'', an open-sourced, graphics processing unit (GPU) enabled Gross--Pitaevskii equation solver \cite{GPUEDOI}. This software allows us to integrate linear and non-linear Schr\"odinger systems in significantly shorter times than alternative implementations \cite{AO:Morgan_pra_2013,Wittekblog_2016}.

To quantify the order of the vortex lattice the position of each vortex was found by summing the wavefunction phase over adjacent grid sites, and looking for a $2\pi$ winding. This gave a vortex position estimated to the numerical grid. A linear least-squares fit was then performed to more accurately determine the vortex core position to sub-grid resolution by locating the real and imaginary zeros of the wavefunction within this region. This allows for the detected core position to take on a continuous range of spatial values within the condensate. As tracking many-body dynamics is a complex problem, we make use of the Delaunay triangulation technique from computational geometry to examine the ordering of the the vortex lattice. In the ideal triangular Abrikosov lattice, every vortex has 6 nearest neighbors, $l=(1,\ldots,6)$, located at $\theta_l=l\pi/3$ around the polar angle and any perturbation related to a defect changes these locations. Delaunay triangulation generates a mesh from the vortex positions, which makes it easy to check for the presence of non 6-fold connected vortices. These vortices are termed as $n$-fold topological lattice defects, where $n$ is the number of connected edges, and dislocation defects can form when, for example, a 5-fold and a 7-fold defect pairs. Since all these structures are easily countable, this method allows us to characterize the effect a well-defined perturbation has on the lattice.

As unperturbed Abrikosov lattices in BECs are well ordered everywhere in the bulk region \cite{Vtx:Anglin_arxiv_2002} we define a radial boundary at approximately $2/3$ of the maximum density, which corresponds to $r_v=2\times 10^{-4}$ m from the center and restrict our analysis to vortices inside it. This leaves an edge boundary of approximately $4a_v$ wide where vortices are not counted. Given the coordinate locations for each vortex within the boundary, it is possible to calculate statistical quantities that characterised the degree to which the lattice is ordered. As our system is of finite size, the usually used translational correlations have only limited value and we will focus in the following on orientational correlations, which quantify how the lattice aligns along a particular angle. The orientational correlation function is defined as
\begin{align}
	g_6(r) = \frac{1}{N(r)}\displaystyle\sum\limits_{j,k}^{N(r)}\zeta_6(\mathbf{r}_j)\zeta_6^{*}(\mathbf{r}_k),
\end{align}
with
\begin{align}
	\zeta_6(\mathbf{r}_{j}) =  \frac{1}{n_j}\displaystyle\sum\limits_{l}^{n_j}\exp(\mathrm{i}6\theta_{jl}),
\end{align}
where $N(r)$ is the number of paired vortices separated by $r=|\mathbf{r}_j - \mathbf{r}_k|$, $\zeta_6$ is the orientational order parameter, $l$ runs over the nearest neighboring vortices, $n_j$ is the number of nearest neighboring vortices, and $\theta_{jl}$ is the angle a paired vortex and a nearest neighbor makes relative to a reference axis \cite{Guillamon_nat_2014}. We examine the orientational correlation function as a measure of the order of a ``vortex unit cell'', defined by the angle made by nearest neighbors to an individual vortex. For a perfectly ordered triangular lattice this value will tend to 1 at $r = a_v$, next nearest-neighbor positions and higher order lattice spacings, and 0 elsewhere.

%%%%%%%%%%%%%%%%%%%%%%%%%%%%%%%%%%%%%%%%%%%%%%%%%%%%%%%%%%%%%%%%%%%%%%%%%%%%%%%%%%%%%%%%%%%%%%%%%%%%%%%%%%%%%%%%%%%%%%%%%%%%%%%%%%%%%%%%%%%%%
\section{Phase imprinting defects}\label{sec:phase}
%%%%%%%%%%%%%%%%%%%%%%%%%%%%%%%%%%%%%%%%%%%%%%%%%%%%%%%%%%%%%%%%%%%%%%%%%%%%%%%%%%%%%%%%%%%%%%%%%%%%%%%%%%%%%%%%%%%%%%%%%%%%%%%%%%%%%%%%%%%%%

Phase imprinting is a class of techniques for directly manipulating the phase of a condensate in such a way that the phase is modified to a desired form \cite{Vtx:Dobrek_pra_1999}. As a consequence the density distribution will adjust itself, and in ground state condensates dark solitons \cite{BEC:Denschlag_science_2000}, as well as vortices \cite{Vtx:Leanhardt_prl_2002,VTX:Brachmann_osa_2011} have been created this way. However, the phase imprinting methods can also be used to annihilate a vortex from the lattice by applying a phase profile of opposite winding to remove the vortex phase singularity.  This will leave the condensate with a density depletion at the prior location of the singularity, which will consequently fill in and excite phonon modes in the condensate.

\subsection{Single vortex dynamics}

\begin{figure}[tb]
    \includegraphics[width=0.45\textwidth]{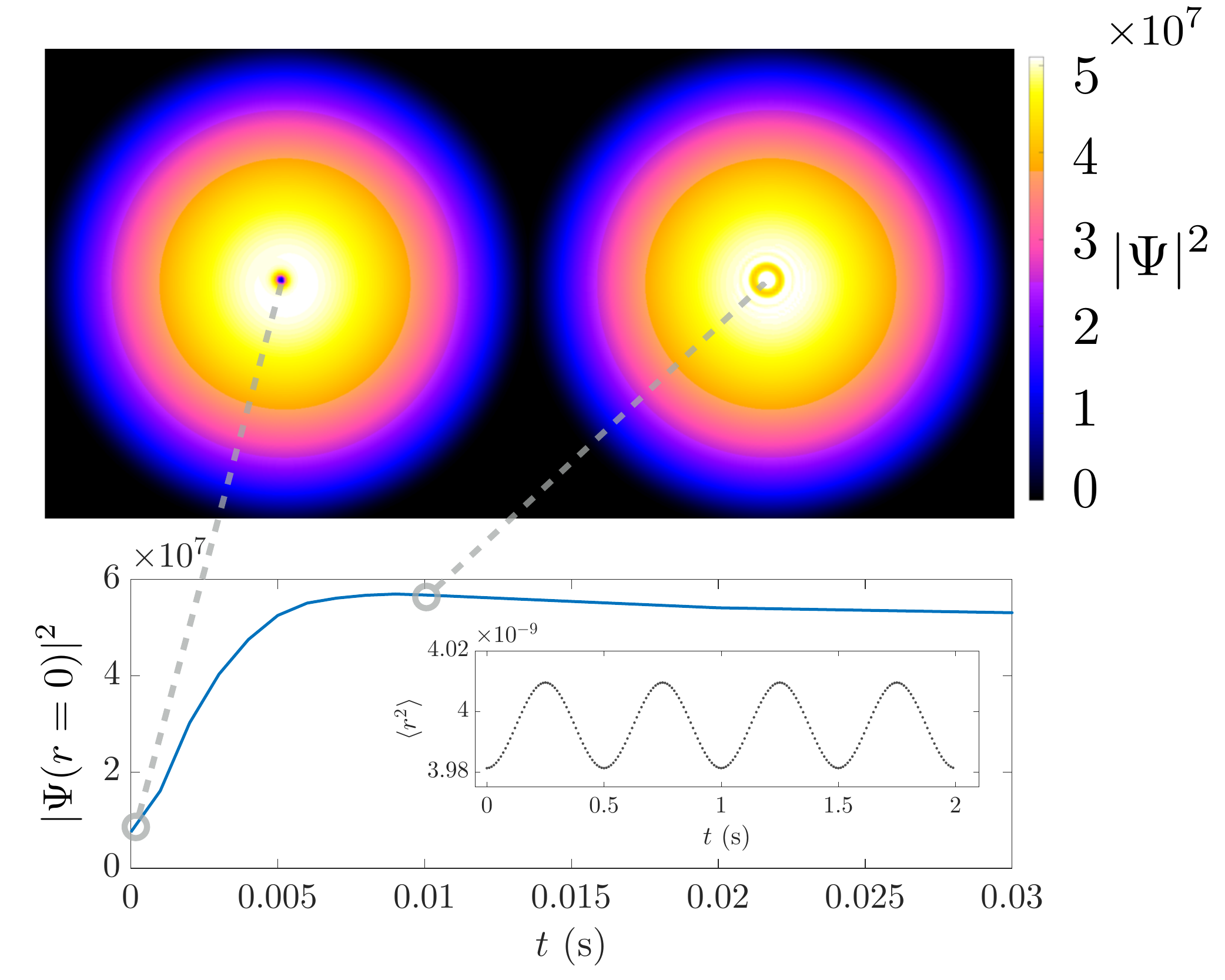}
    \caption{The evolution of the condensate density is shown for the initial state, and after 10 ms of evolution. The removal of the phase singularity at $r=0$ leads to a filling in of the density dip, which can also be seen from the line-plot. The process excites the monopole mode at frequency $2\omega_\perp$ (see inset).}\label{fig:annihilation_1vtx}
\end{figure}

\begin{figure}[tb]
    \includegraphics[width=0.35\textwidth]{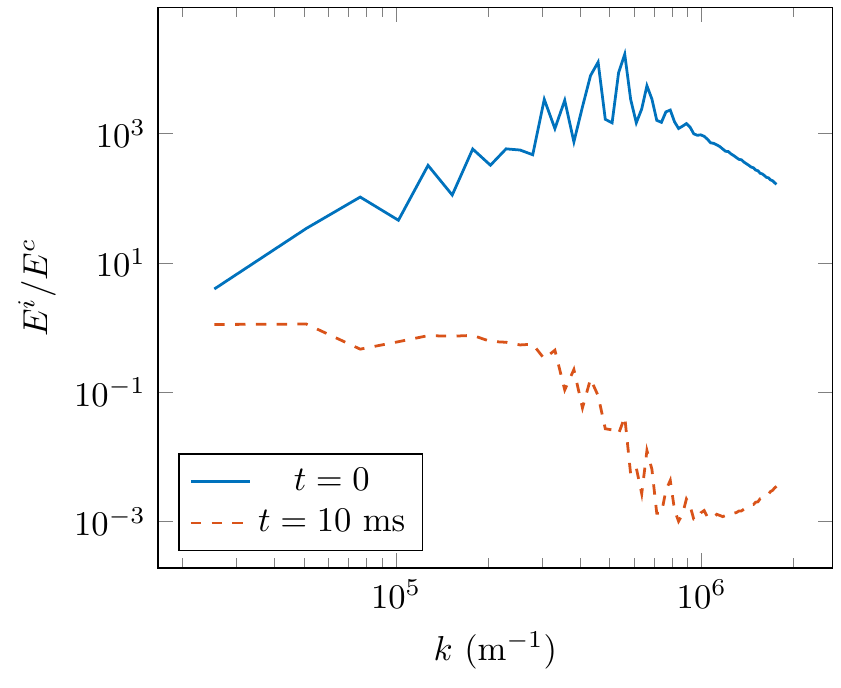}
    \caption{Ratio of incompressible to compressible energy at  $t=0$ (solid) and $t=10$ ms (dashed). Initially the incompressible energy is greater than the compressible due to the presence of the vortex, giving values greater than unity for all $k$. After application of the phase profile, the vortex is annihilated, with the energy released as phonons, indicated by a decrease in incompressible energy for all $k$ values.}\label{fig:kinspec}
\end{figure}

To fully understand the effects of removing a vortex from the lattice system, let us first investigate the situation where the vorticity from a condensate carrying only a single vortex is removed. For this we apply a phase pattern that exactly cancels the $2\pi$ phase winding and simulate the resulting dynamics. The results of such a process can be seen in Fig.~\ref{fig:annihilation_1vtx}, and, as expected, the depletion in the condensate density fills in after the vortex phase is removed and the breathing mode is excited. Since the system is rotationally symmetric, we also plot the expectation value of the squared radius, $\langle r^2 \rangle$, where $r^2 = x^2 + y^2$, which clearly shows that the annihilation process excites the breathing mode at the expected frequency of $2\omega_\perp$ for a two-dimensional system \cite{BEC:Pitaevskii_pra_1997}. The change in energy due to the phase removal can be meaningfully characterised via the ratio of compressible (phonon) and incompressible (vortex) kinetic energy spectra shown in Fig.~\ref{fig:kinspec}. The kinetic energies are determined from the density weighted velocity field, $\mathbf{u} = |\Psi|\frac{\hbar}{m}\nabla \theta$, where $\theta$ is the phase of the condensate. The contribution from the compressible and incompressible energies can be taken as $\mathbf{u} = \mathbf{u}^{c} + \mathbf{u}^{i}$, and are determined by solving
\begin{subequations}
    \begin{align}
        \nabla \times \mathbf{u}^{c} = 0 \\
        \nabla \cdot \mathbf{u}^{i} = 0.
    \end{align}
\end{subequations}
The resulting spectra, $E^{i,c}$, are then calculated as an angle average over linearly spaced intervals of wavenumber magnitudes in reciprocal space \cite{VTX:Bradley_prx_2012}
\begin{equation}
    E^{i,c}(k) = \frac{mk}{2}\displaystyle\sum_{j\in r}\int\limits_{0}^{2\pi} d\phi_k \frac{\mathcal{U}^{i,c}_j(\mathbf{k},t)}{s_k},
\end{equation}
where
\begin{equation}
    \mathcal{U}^{i,c}_{j}(\mathbf{k},t) = \int d^2\mathbf{r}e^{-\mathrm{i}(\mathbf{k}\cdot \mathbf{r})}u_j^{i,c}(\mathbf{r},t),
\end{equation}
and $s_k$ is the number of samples in a particular interval. As one can see from Fig.~\ref{fig:kinspec} after the vortex is annihilated and sounds wave are created, the energy ratio drops and lower incompressible-to-compressible values appear, in particular for higher wavenumbers. The latter is due to the removal of large kinetic energies from the atoms close to the vortex core.


\begin{figure}[tb]
    \includegraphics[width=0.45\textwidth]{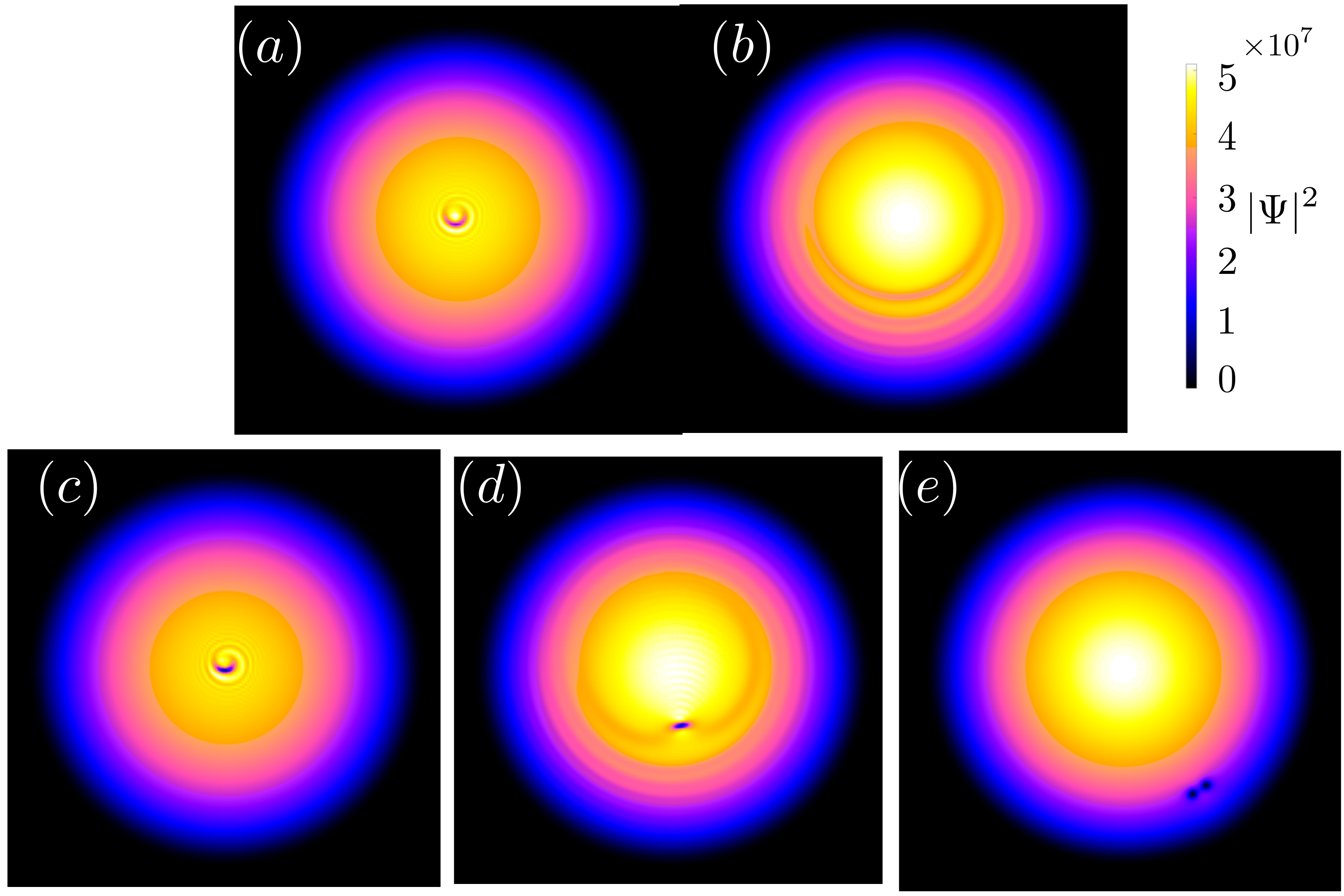}
    \caption{The condensate evolution following an uncentered phase imprint. For an imprint where the singularities of the vortex and the imprinted phase are less than twice the healing length away from each other the existing vortex is annihilated and phonon modes are excited $(a,b)$. However, beyond this distance an antivortex is created, which travels with the pre-existing vortex and circulates the condensate $(c,d,e)$. The distance for cases $(a,b)$ and $(c,d,e)$ are $r = 1.36\times10^{-6}$ m, and $r =2.73 \times10^{-6}$ m respectively.}\label{fig:annihilation_1vtx_uncentred}
\end{figure}

While the above example suggests that erasing vortices is a straightforward and controllable process, this assumption needs to be checked for the situation where the imprinted phase and the existing phase are not perfectly centered on each other. This situation is shown in Fig.~\ref{fig:annihilation_1vtx_uncentred}, and one finds that cases where the imprinted profile is sufficiently close to the core (i.e. within twice the healing length, $\xi\approx1.06\times 10^{-6}$ m) the existing vortex gets erased as before. However, beyond this distance a separate antivortex gets created and the vortex-antivortex pair travels to the edge of the condensate system and begins to circulate around \cite{VTX:Martikainen_pra_2001}. For a densely packed lattice of vortices, however, this is not a problem since the typical distance between vortices is of the same order of magnitude as the healing length.

\subsection{Lattice dynamics}

%%%%%%%%%%%%%%%%%%%%%%%%%%%%%%%%%%%%%%%%%%%%%%%%%%%%%%%%%%%%%%%%%%%%%%%%%%%%%%%%%%%%%%%%%%%%%%%%%%%%%%%%%%%%%%%%%%%%%%%%%%%%%%%%%%%%%%%%%%%%%

The removal of a single vortex from the vortex lattice by phase erasing initially affects only the nearest neighbors, as the phase gradient is only significant over the length scale of a healing length close to the erased singularity. The altered velocity profile will lead to the remaining vortices leaving their position in the Abrikosov lattice and the excitation of phonon modes. However in the lattice areas away from the impurity, these phonon modes have only minimal impact on the geometry \cite{VTX:oriordan_pra_2016}. To characterize the vortex dynamics following the application of the phase profile, we will in the following track each individual vortex throughout the full time-evolution and use the resulting trajectories and Delaunay triangulation for analysis.

\begin{figure}
\includegraphics[width=0.48\textwidth]{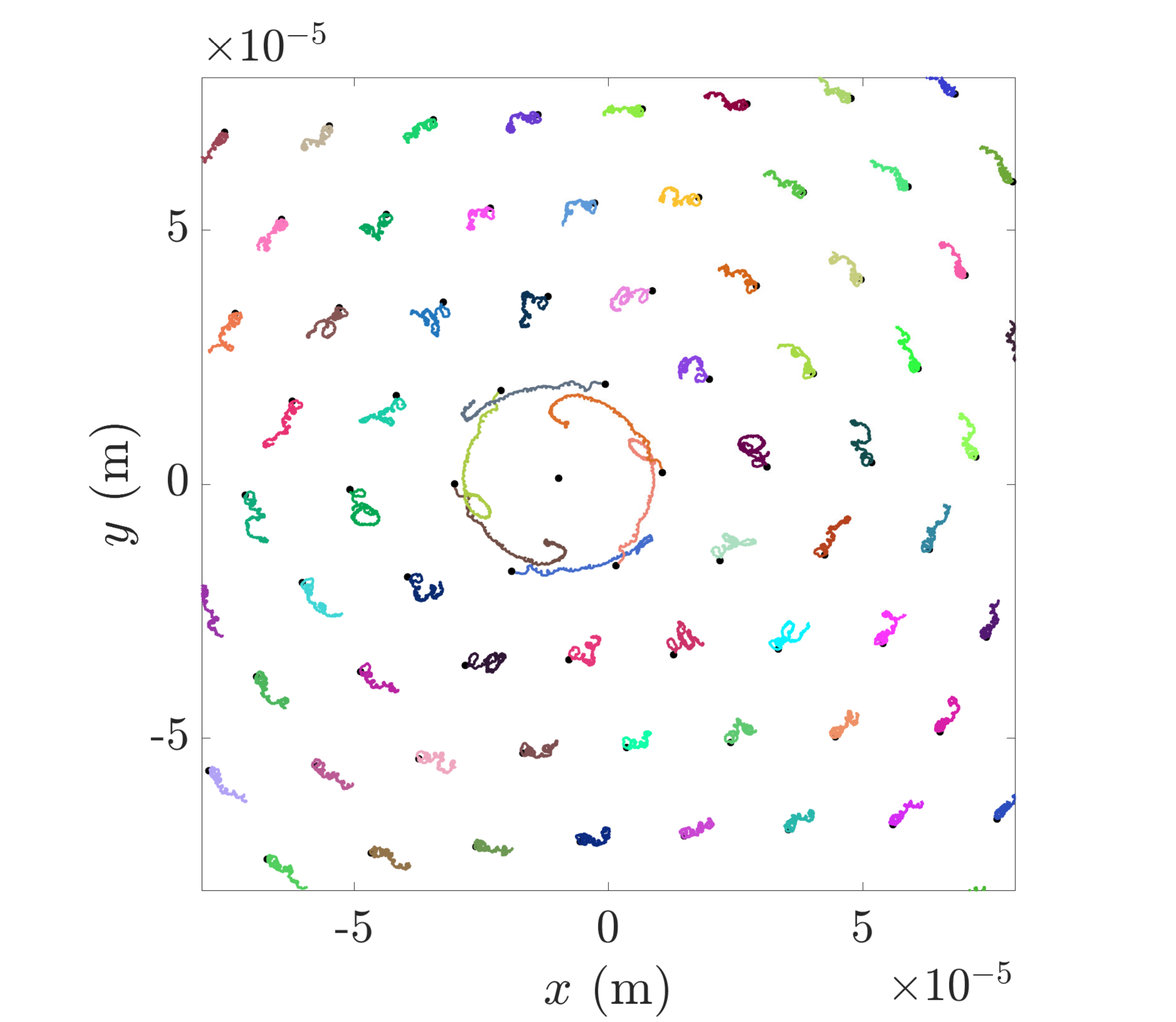}
    \caption{The trajectories of the vortices over 4 seconds following the removal of the vortex closest to the center, where each color represents a unique trajectory path. The vortices can be seen to move counter-clockwise in the co-rotating frame due to the loss of the local velocity field. However, the effect of the removal decreases quickly with increased radial distance. }
    \label{fig:trajplot}
\end{figure}

\begin{figure}[tb]
    \includegraphics[width=0.48\textwidth]{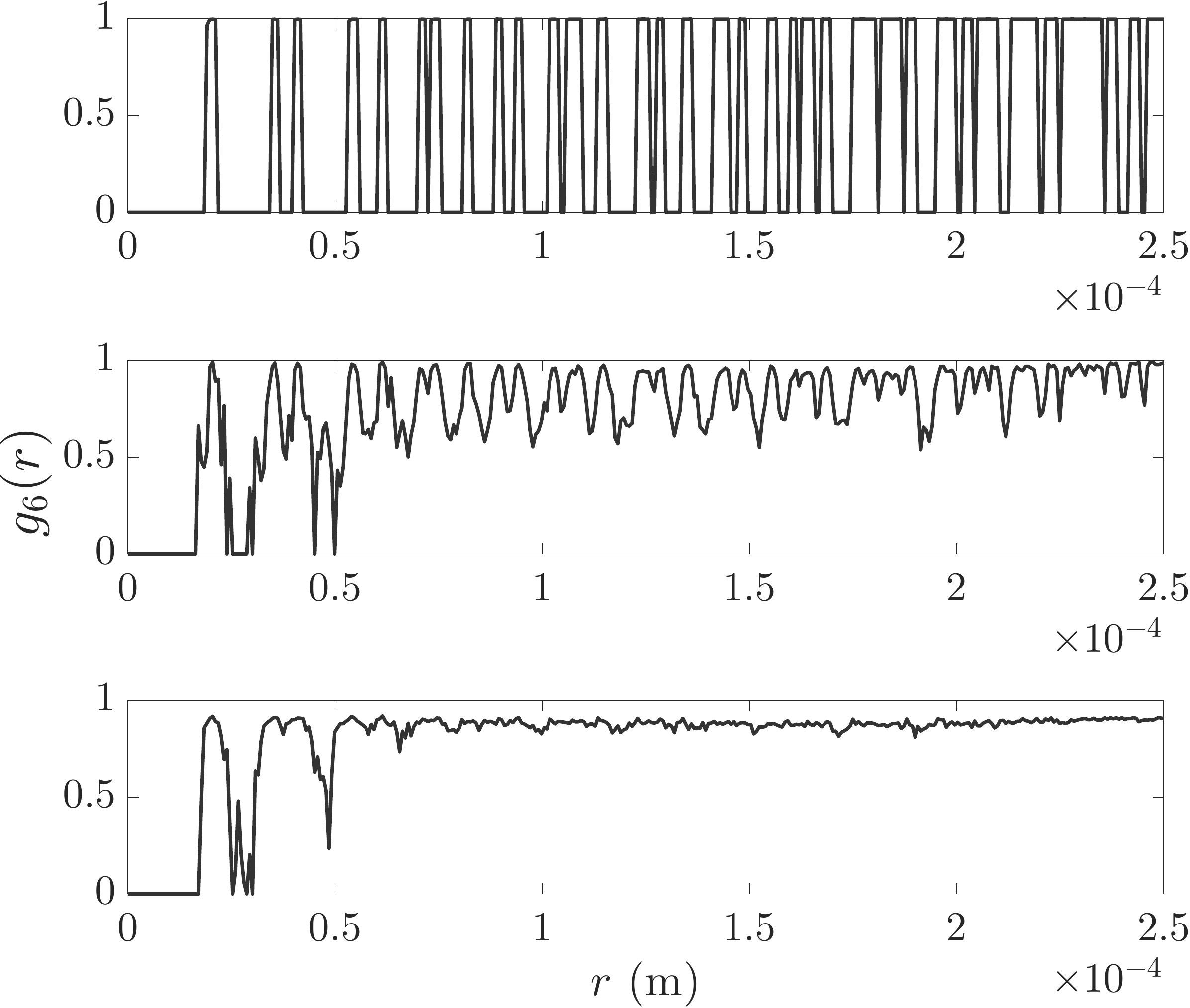}
    \caption{The orientational correlation function for the vortex lattice after removing the central vortex is given for $t=(0,1,6)$ seconds (top to bottom). The peaks at $t=0$ appear at nearest neighbor, next nearest, and higher order distances. Due to finite binning of the lengths, the peaks become grouped to 1 at higher length scales. For times greater than $t=0$, the peak correlations drop, however, the large value at long times indicates a well ordered lattice as high correlations are observed across all length scales.}\label{fig:g6}
\end{figure}

Let us first consider the situation where a single vortex is erased within the central area of the vortex lattice. In Fig.~\ref{fig:trajplot} we show the trajectories of the remaining vortices over a time-scale of 4 seconds. One can see that a long-lived vacancy is maintained close to the centre with the adjacent vortices rotating faster than the lattice due to the loss of the local velocity field. The honeycomb-like vacancy region eventually decays and the system settles into a new local geometry. Very similar behavior can be observed if the erased vortex is not one of the central ones, as long as it is within a region of constant areal vortex density. However, being closer to the edge of the lattice reduces the stability of the perturbed region. The overall lattice remains well structured after a vortex removal, as can be seen from the orientational correlation function shown in Fig.~\ref{fig:g6} for different times. Although the gaps between the peaks that exist at $t=0$ disappear during the evolution due to the presence of the phonon excitation, the overall correlations remain high for long times and constant across all length scales. The slight peak softening arises from the vortices no longer being aligned to a perfect triangular lattice position, which is indicative of a weak disordering or distortion of the lattice structure.

\begin{figure}[bt]
    \includegraphics[width=0.48\textwidth]{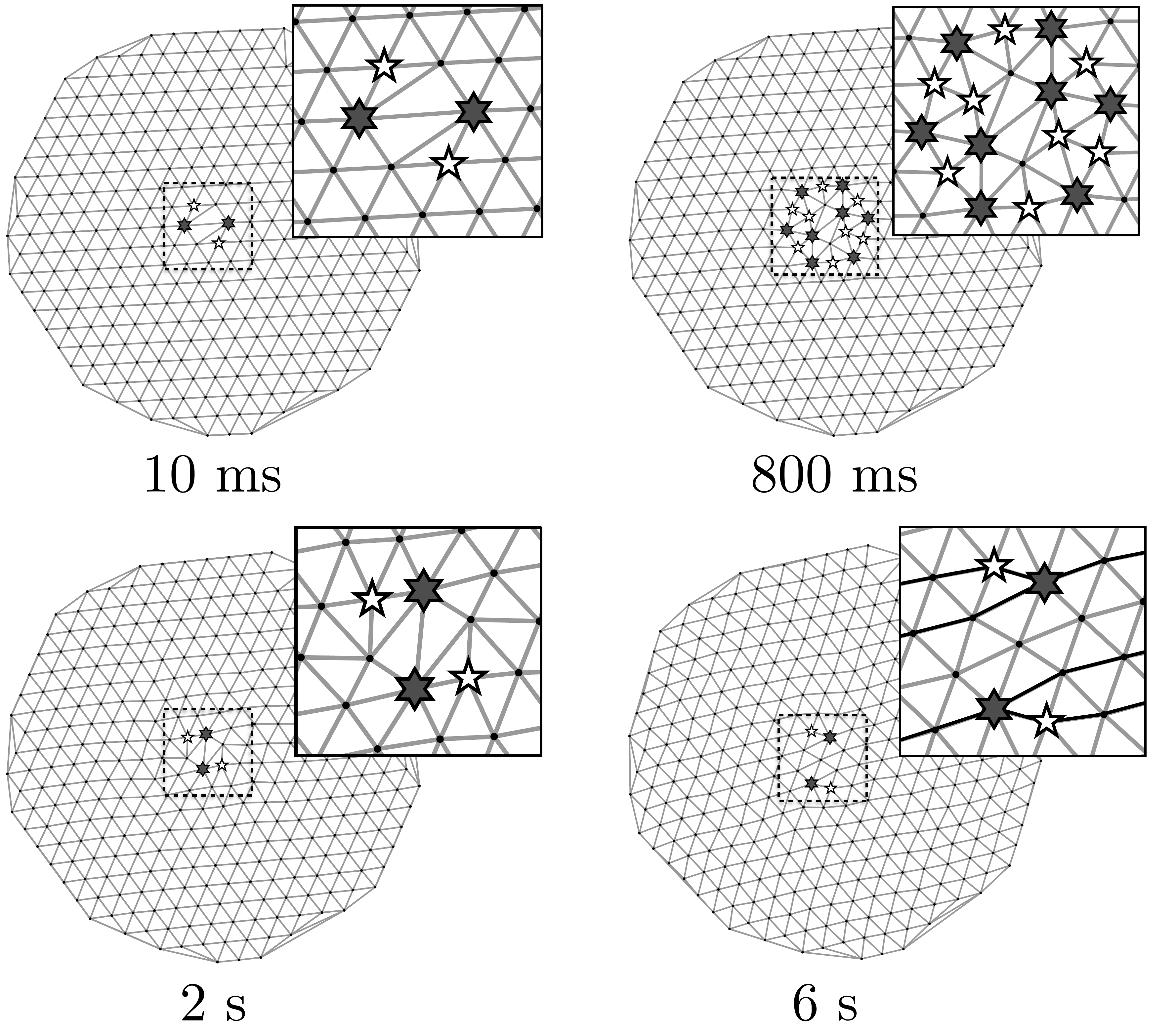}
    \caption{Delaunay triangulation of the vortex lattice after removing one vortex, shown at $t=(0.01,0.8,2,6)$ s. The resulting lattice defects are indicated by white and gray stars for 5-fold and 7-fold defects respectively. One can see that two (5,7) dislocations are formed quickly, which settle and persist in the lattice for long times. Lattice dislocation lines are indicated for inset $t=6$ s.}\label{fig:deltri_1vtx}
\end{figure}

As described above, the Delaunay triangulation of the lattice can give a graphical overview of how connected the different vortices are, and therefore what changes to the lattice structure have occurred \cite{Guillamon_nat_2014}. We show the resulting graph for the case where a single vortex was removed from the center of the lattice in Fig.~\ref{fig:deltri_1vtx}. One can see that a pair of (5,7)-fold connected lattice defects form immediately after the removal (at 10 ms), which slightly adjusts and becomes stable for long times. Removing vortices at different positions in the lattice shows similar behavior, with a localization of the disordered region not far from the site of vortex removal.

\begin{figure}[tb]
    \includegraphics[width=0.48\textwidth]{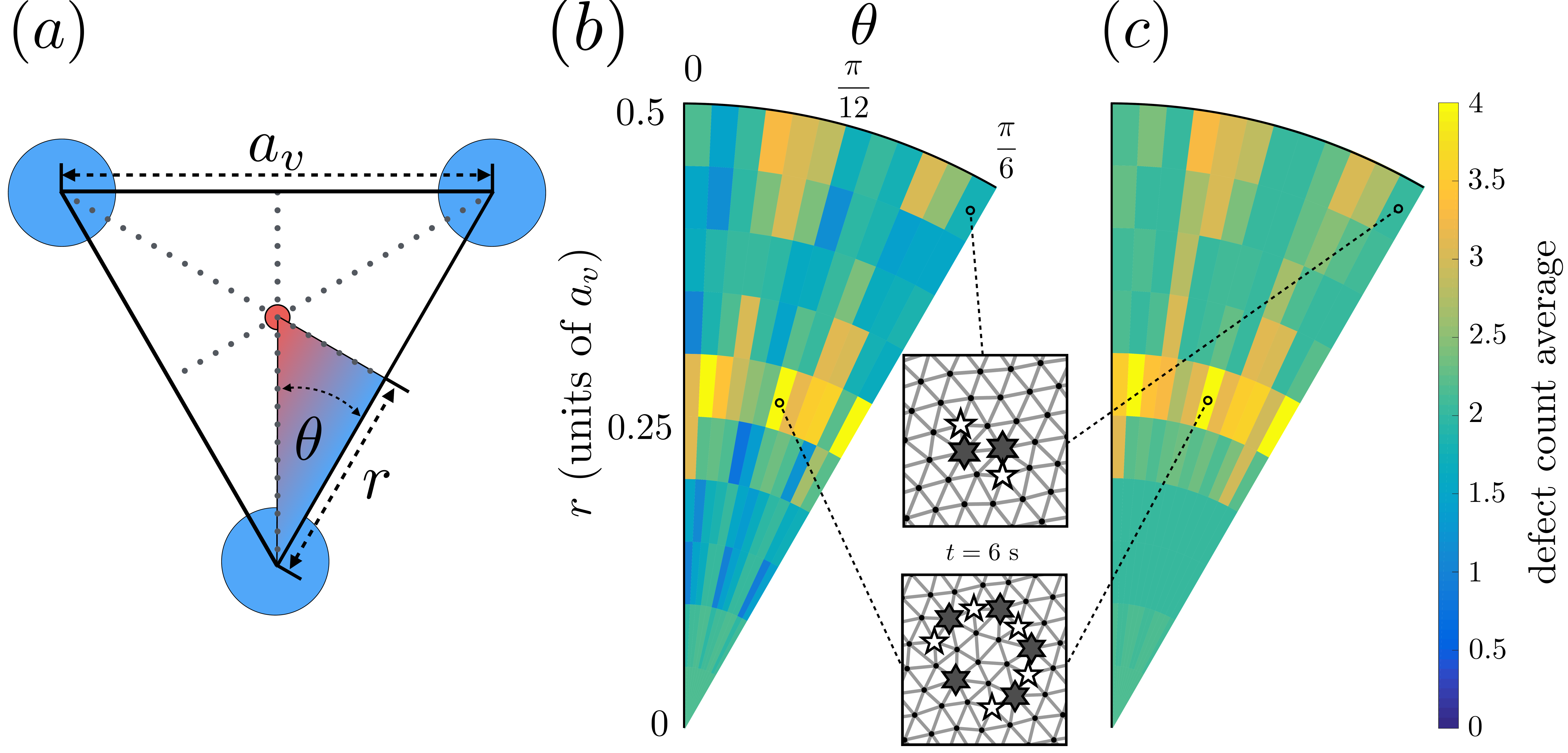}
    \caption{The time-averaged number of defects appearing over a range of imprint positions, relative to a central vortex from $t=1\rightarrow$10 s, and allowing 1 s of settling time. A schematic of the examined region is shaded in (a), with the resulting 5-fold ($b$) and 7-fold ($c$) defects shown following an imprint. The insets show a snapshot of two different parameter regions at $t=6$ s. A high simultaneity is observed between their appearance, where a paired (5,7) defect indicates a lattice dislocation. Not all 5 and 7-fold defects pair, as some can exist individually, or pair with other $n$-fold defects.} \label{fig:lattice_misalign}
\end{figure}

If the phase imprinting is not directly aligned with the vortex singularity, other $n$-fold dislocations can be found in the Delaunay triangulation. This is due to the vortex core size becoming comparable to the average spacing between the cores in rapidly rotating condensates, and therefore the imprinted changes to the velocity field affect more vortices. Fig.~\ref{fig:lattice_misalign} shows the time-averaged number of lattice defects following an imprint displaced relative to a vortex and lattice vector. One can see that if the displacement is still within the core of the vortex, on average 1 or 2 defects are created of the 5-fold ($a$) and the 7-fold ($b$) kind. At the cusp of the core, the imprint tends to create upwards of 3 to 4 defects, which again tends back to the average of 2 beyond this region. This shows that the previously discussed issue resulting from the creation of antivortices through imperfect alignment does not exist in Abrikosov lattices, and we will concentrate on the perfect imprint of the phase in the following discussions.

\begin{figure}[tb]
    \includegraphics[width=0.48\textwidth]{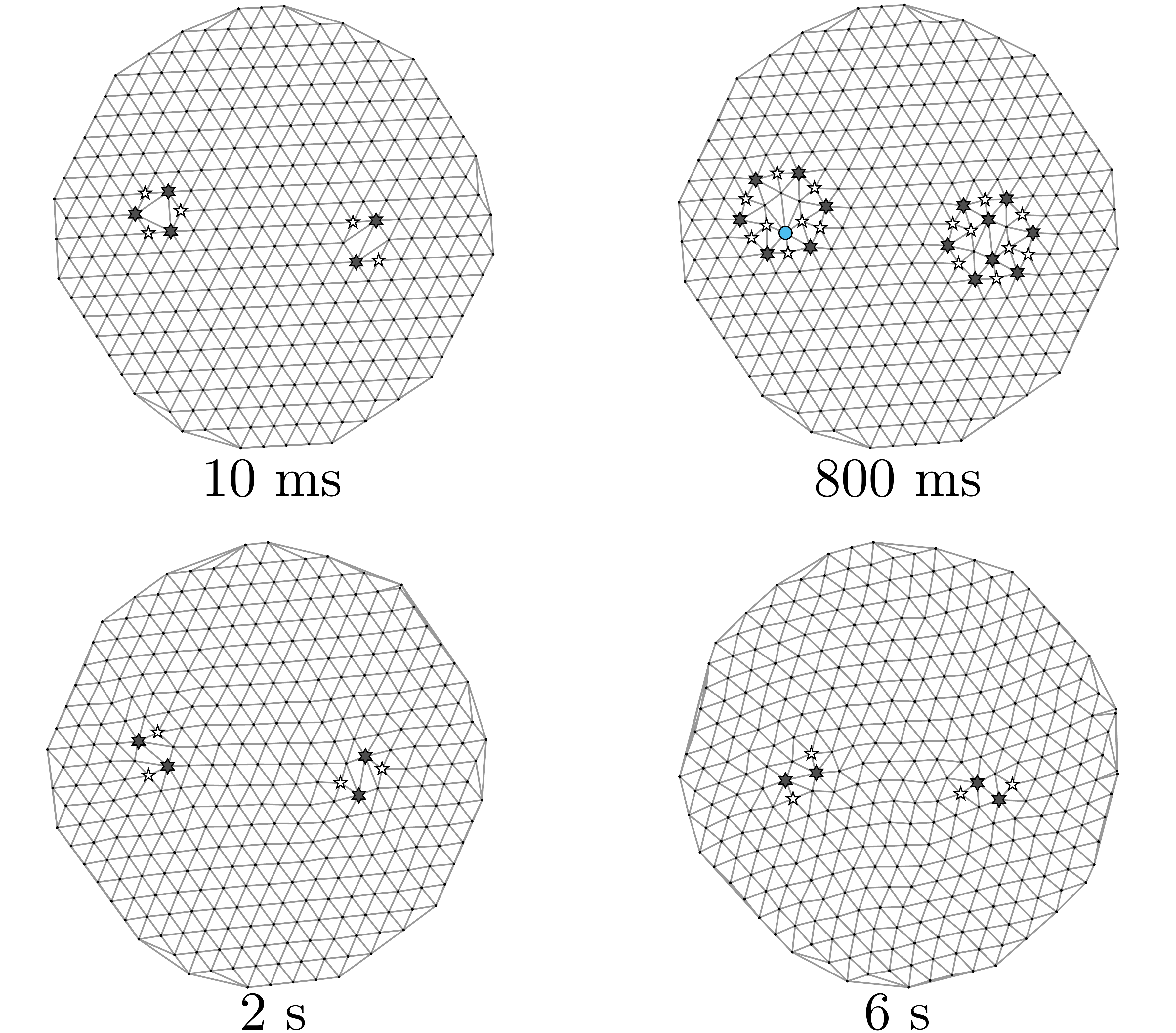}
    \caption{Delaunay triangulation of the vortex lattice upon removal of two vortices at either sides of the lattice for $t=(0.01,0.8,2,6)$ s. The resulting defects that form remain localized for long times. The lattice largely remains ordered, as observed with removing the central vortex.}\label{fig:traj_2vtx_edge}
\end{figure}

\begin{figure}[bt]
    \includegraphics[width=0.48\textwidth]{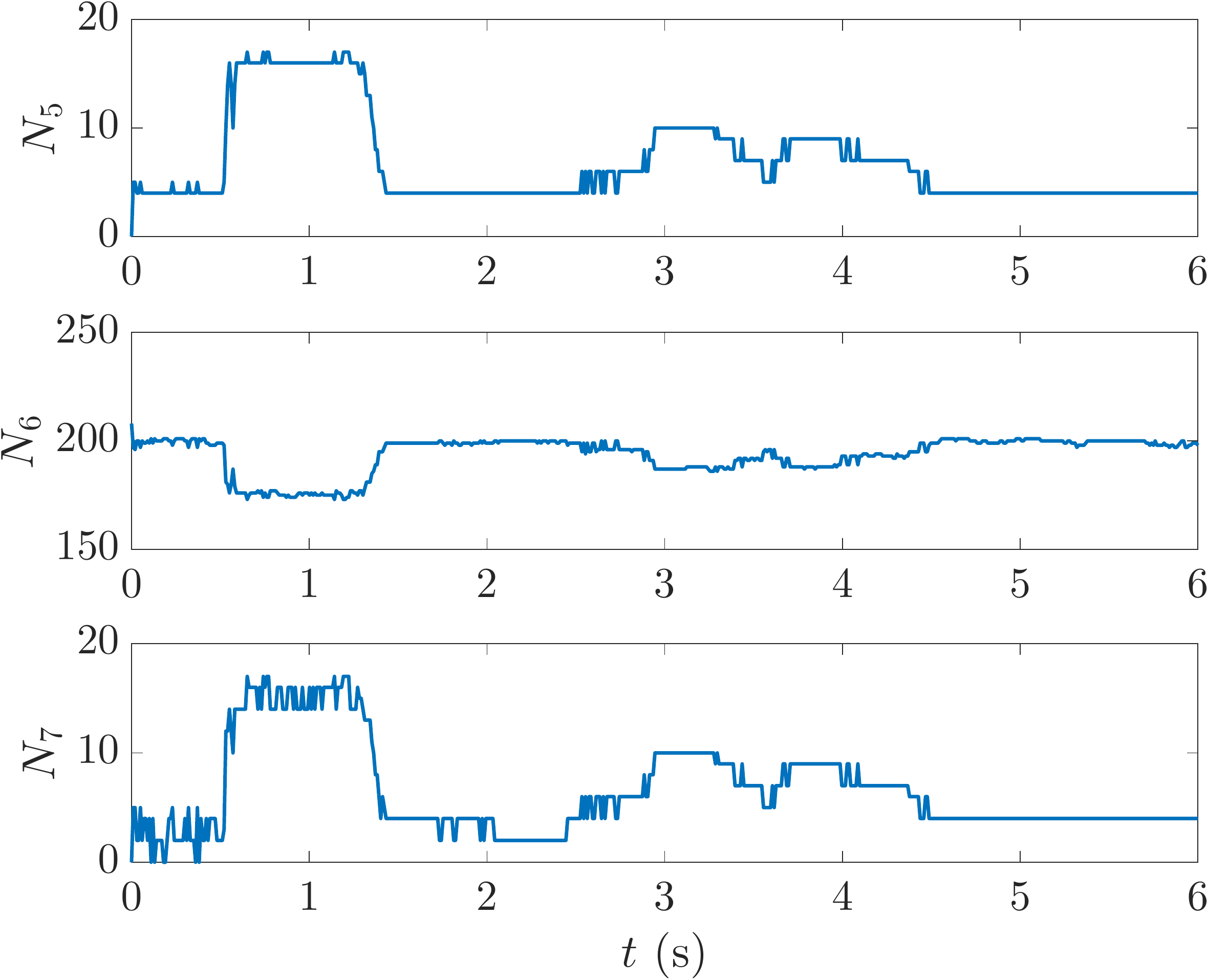}
    \caption{The defect count taken from a Delaunay triangulation of the vortex lattice following the removal of two vortices on opposite sides as a function of time. After a brief settling time, the lattice attains an almost constant defect count.}\label{fig:remove7_defect}
    \label{fig:vtx_rem2_edge}
\end{figure}

To further demonstrate the localized nature of the defect, let us briefly discuss the situation where two vortices are erased in separate regions away from the lattice centre. The Delaunay triangulation for this case is shown in Fig.~\ref{fig:traj_2vtx_edge}, and the independence of the two localized regions is clearly visible, with each  showing similar behavior to the case discussed above. Since we are limiting ourselves here to perfect imprinting, we also show the number of edges formed between vortices as a function of time for 5, 6 and 7 nearest neighbors respectively ($N_x$) in Fig.~\ref{fig:vtx_rem2_edge}. One can see that the initial perturbation settles quickly to values similar to the ones above.

%%%%%%%%%%%%%%%%%%%%%%%%%%%%%%%%%%%%%%%%%%%%%%%%%%%%%%%%%%%%%%%%%%%%%%%%%%%%%%%%%%%%%%%%%%%%%%%%%%%%%%%%%%%%%%%%%%%%%%%%%%%%%%%%%%%%%%%%%%%%%
%%%%%%%%%%%%%%%%%%%%%%%%%%%%%%%%%%%%%%%%%%%%%%%%%%%%%%%%%%%%%%%%%%%%%%%%%%%%%%%%%%%%%%%%%%%%%%%%%%%%%%%%%%%%%%%%%%%%%%%%%%%%%%%%%%%%%%%%%%%%%

\begin{figure}[tb]
    \vspace{1cm}
    \includegraphics[width=0.48\textwidth]{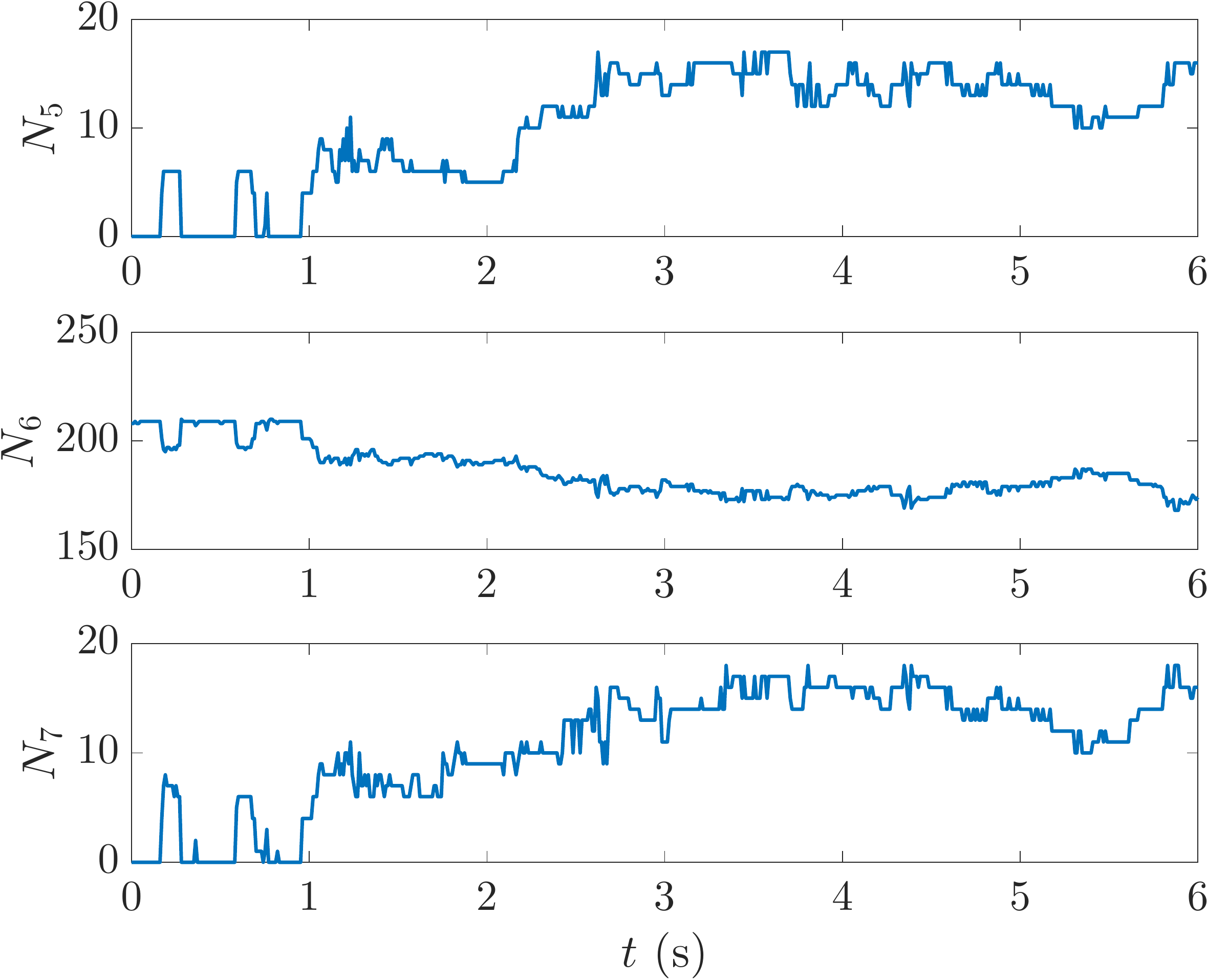}
    \caption{The defect count taken from a Delaunay triangulation of the vortex lattice following an insertion of an antivortex. The number of defects increases as the local structure decays, and eventually gives rise to a quasi-constant state.}\label{fig:varr161anti_defect}
\end{figure}

In addition to simply erasing vorticity, we can also use phase imprinting to create varying degrees of disorder. By, for example, applying an appropriate $4\pi$ magnitude phase imprint we can replace a vortex with an antivortex at a given position. Since this does not require a change in the local density, all resulting perturbations stem from the adjusted velocity field of the vortex that has been flipped \cite{VTX:Madarassy_gfd_2009}. However, it is immediately obvious that such a situation is unstable, which can be confirmed by observing the creation of a large number of defects, as shown in Fig.~\ref{fig:varr161anti_defect}. An increase in the number of defects can be seen up to approximately $t=3$ s, during which the antivortex causes local disordering of the lattice, annihilates with a nearby vortex, and gives rise to the creation of a large number of (5,7) defect pairs. After this the number of defects no longer grows, but instead fluctuates about a stable value which is greater that that of the previously examined cases.

A final class of possible perturbations is the removal of a cluster of neighboring vortices from the lattice, and in Fig.~\ref{fig:remove7_defect} we show the results from erasing an entire seven vortex unit cell from the condensate. As expected, one can see that the number of lattice defects rises considerably and does not settle during the time over which we can simulate the condensate. In this case, the disordered regions occupy a large area of the lattice and the number of 6-fold connected vortices becomes very low.

\begin{figure}[tb]
    \includegraphics[width=0.48\textwidth]{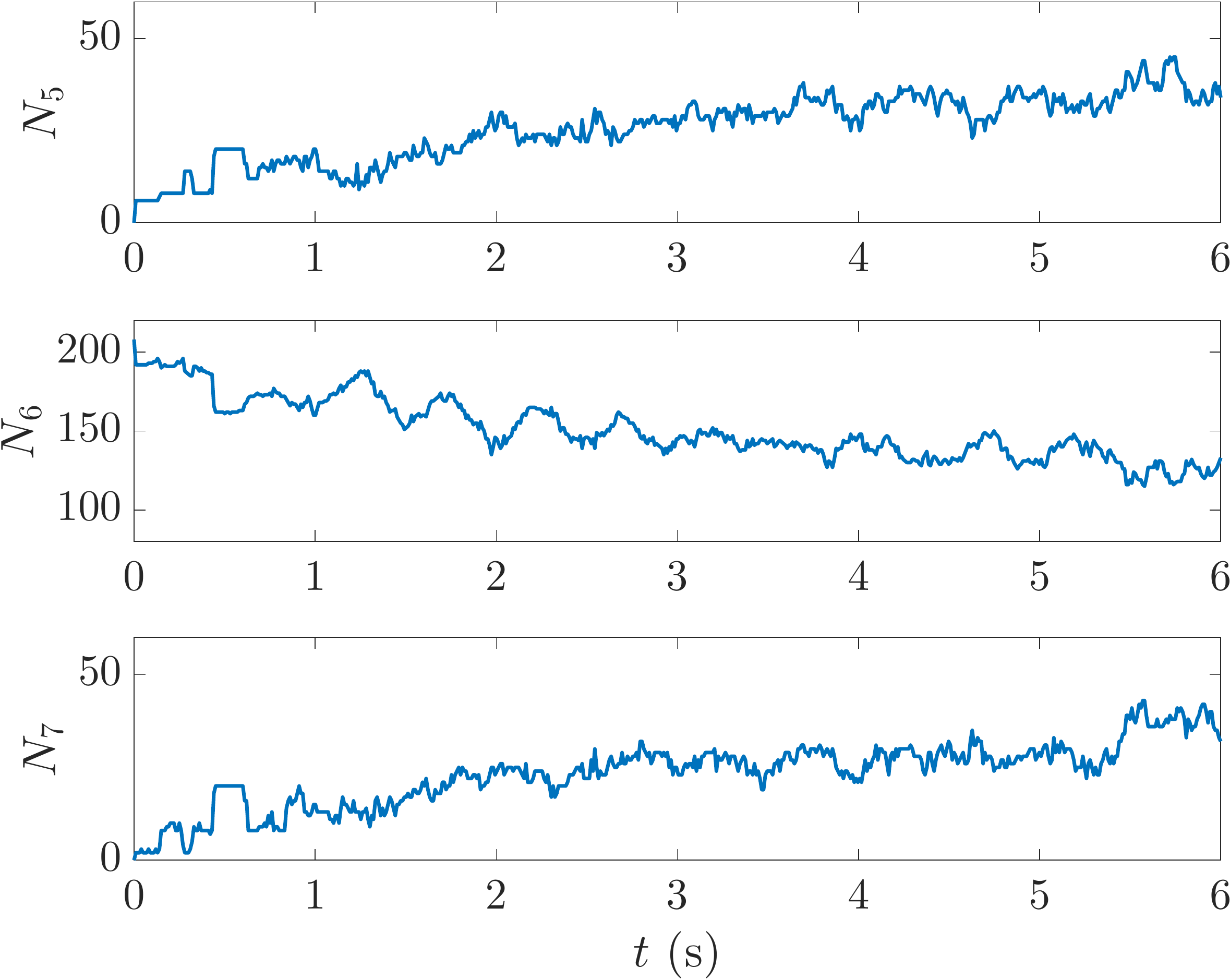}
    \caption{The defect count taken from a Delaunay triangulation of the vortex lattice following the removal of 7 vortices from the centre of the lattice. }\label{fig:remove7_defect}
\end{figure}

\begin{figure}[bt]
    \includegraphics[width=0.48\textwidth]{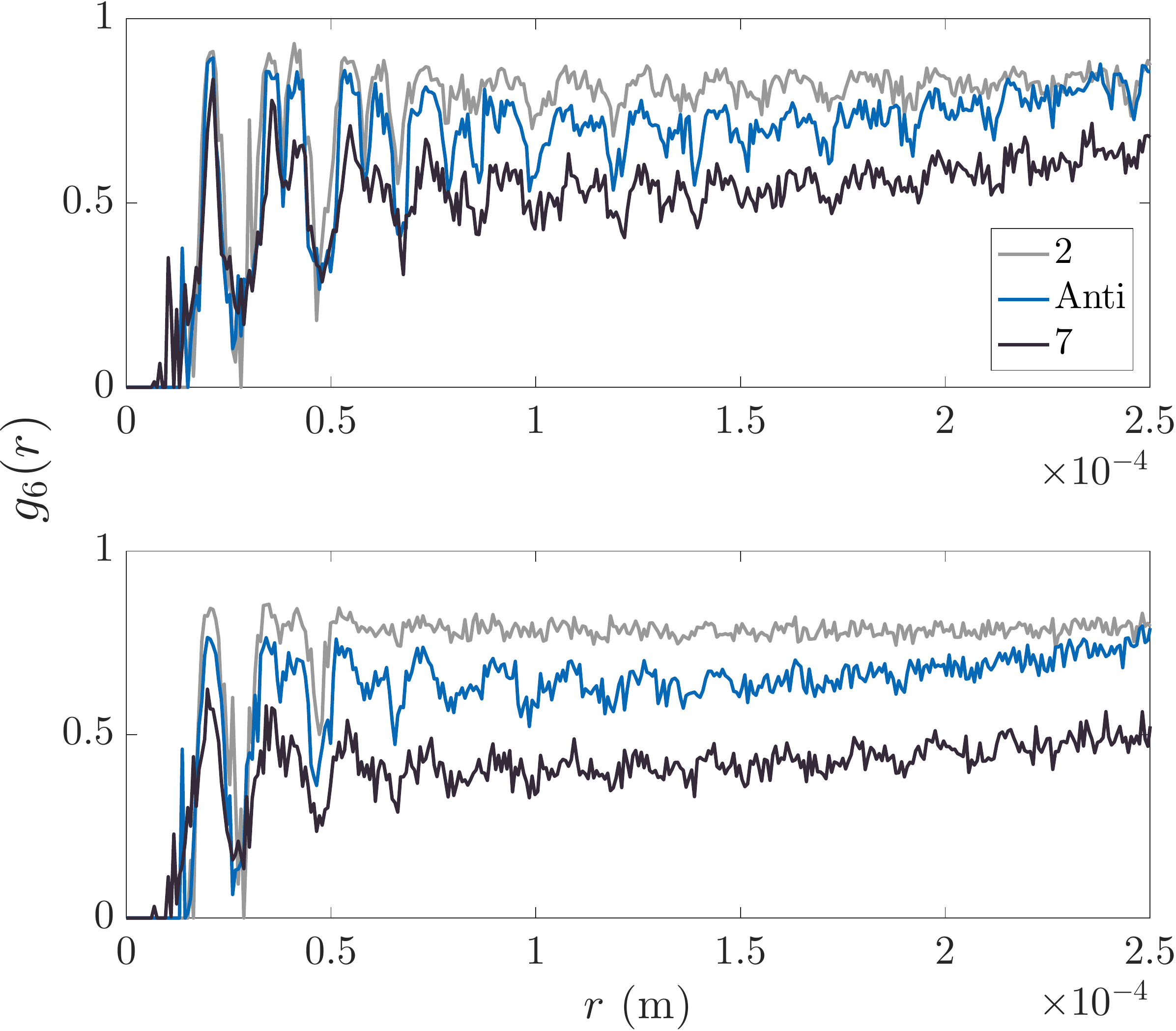}
    \caption{The orientational correlation function is given for moderate ($t=3$ seconds, top) and long ($t=6$ seconds, bottom) times after the phase imprint. The general behaviour at short and long ranges is similar for all three scenarios, but the correlations are significantly reduced, especially for the situation where seven vortices are removed.}\label{fig:g6_2edge_anti_nuclear}
\end{figure}

Comparing the orientational correlation functions for the three cases discussed in this section (removing two distant vortices, creating an antivortex, and removing 7 vortices) also demonstrates the different degree of disorder they produce (see Fig.~\ref{fig:g6_2edge_anti_nuclear}). The removal of the two vortices at opposite sides of the condensate still yields reasonably high correlations ($\langle g_6(r) \rangle \approx 0.8$) at all times and length scales, indicating a well ordered lattice. Creating an antivortex in the lattice leads to lower correlations across all length scales, especially in the long time limit ($\langle g_6(r) \rangle\approx 0.7$), but still tends to the same long-ranged value as the previous case. This indicates an ordered lattice outside the region of the localized defects. Lastly, the removal of seven vortices shows a significant drop in correlations at all length scales and across both times ($\langle g_6(r) \rangle\approx 0.5$), indicating a global disordering of the vortices, which is consistent with the large number of defects identified earlier.

%%%%%%%%%%%%%%%%%%%%%%%%%%%%%%%%%%%%%%%%%%%%%%%%%%%%%%%%%%%%%%%%%%%%%%%%%%%%%%%%%%%%%%%%%%%%%%%%%%%%%%%%%%%%%%%%%%%%%%%%%%%%%%%%%%%%%%%%%%%%%
\section{Discussion and outlook}\label{sec:Conclusions}
%%%%%%%%%%%%%%%%%%%%%%%%%%%%%%%%%%%%%%%%%%%%%%%%%%%%%%%%%%%%%%%%%%%%%%%%%%%%%%%%%%%%%%%%%%%%%%%%%%%%%%%%%%%%%%%%%%%%%%%%%%%%%%%%%%%%%%%%%%%%%

We have shown that the removal of a vortex from an Abrikosov vortex lattice via phase imprinting creates a quasi-stable honeycomb-like vacancy site. The removal of the associated velocity field near the vacancy, however, disturbs the solid-body behavior of the lattice and the vacancy region rotates slower than the surrounding vortex lattice. It eventually decays, creating highly stable topological lattice defects that persist for long times. In fact, the resulting defects can be seen to pair, with (5,7) lattice defects being the most prominent, and manifesting themselves as dislocation defects in the lattice.

The characterization of perturbed lattices put forward by us complements the recent work of Rakonjac \textit{et al.} \cite{Rakonjac:16}, where the authors determine the disorder present in a vortex lattice in a BEC by comparing the ratio of the standard deviation of nearest neighbor distances to the mean distance. Here we extend the available tools by using orientational correlations, Delaunay triangulation for topological defect detection, and by introducing a method to controllably engineer lattice defects through phase imprinting. Although our work relies on theoretical {\it data}, all methods can easily be applied to experimental data as well.

We have also discussed that various kinds of defects can be created controllably by varying the degree to which the crystal structure gets erased. This suggests the use of the phase erasing technique in Abrikosov lattices for examining turbulence in two-dimensional condensates. By introducing a variable number of antivortices one could, for example, examine the transition from an ordered to a disordered system.

\bibliographystyle{unsrt}

\end{document}